# Exotic stable calcium carbides

Yan-Ling Li[1], Sheng-Nan Wang[2], Artem R. Oganov[2-5]
Huiyang Gou[6], Jesse S. Smith[7], Timothy A. Strobel[6]

1 *Laboratory for Quantum Design of Functional Materials, School of Physics and Electronic Engineering, Jiangsu Normal University, 221116, Xuzhou, People's Republic of China*
2 *Department of Geosciences, State University of New York, Stony Brook, NY 11794-2100, USA*
3 *Center for Materials by Design, Institute for Advanced Computational Science, State University of New York, Stony Brook, NY 11794-2100, USA*
4 *Moscow Institute of Physics and Technology, 9 Institutskiy lane, Dolgoprudny city, Moscow Region, 141700, Russian Federation*
5 *Northwestern Polytechnical University, Xi'an, 710072, People's Republic of China*
6 *Geophysical Laboratory, Carnegie Institution of Washington, Washington, DC 20015, USA*
7 *High Pressure Collaborative Access Team, Geophysical Laboratory, Carnegie Institution of Washington, Argonne, Illinois 60439, USA*

**It is well known that pressure causes profound changes in the properties of atoms and chemical bonding, leading to the formation of many unusual materials. Here we systematically explore all stable calcium carbides at pressures from ambient to 100 GPa using variable-composition evolutionary structure predictions. We find that $Ca_5C_2$, $Ca_2C$, $Ca_3C_2$, $CaC$, $Ca_2C_3$, and $CaC_2$ have stability fields on the phase diagram. Among these, $Ca_2C$ and $Ca_2C_3$ are successfully synthesized for the first time via high-pressure experiments with excellent structural correspondence to theoretical predictions. Of particular significance are the base-centered monoclinic phase (space group $C2/m$) of $Ca_2C$, a quasi-two-dimensional metal with layers of negatively charged calcium atoms, and the primitive monoclinic phase (space group $P2_1/c$) of CaC with zigzag $C_4$ groups. Interestingly, strong interstitial charge localization is found in the structure of $R$-$3m$-$Ca_5C_2$ with semimetallic behaviour.**

Unexpected chemical reactions can happen under extreme conditions, with emergence of rich phase diagrams and materials possessing intriguing properties. Recently, by combining variable-composition structure prediction methods with first-principles total energy calculations[1], pressure-composition ($P$-$x$) phase diagrams were predicted for such binary systems as Mg-O[2] and Na-Cl[3]. In both cases, the unexpected compounds that were predicted have been successfully synthesized[3] (For the Mg-O system, Goncharov, A. F., personal communication). Elemental carbon and calcium both exhibit rich diversity of stable and metastable phases under pressure[4-6]. Compressed calcium shows unique structural and electronic properties and exhibits the highest recorded superconducting critical temperature among pure elements[7]. For carbon, only graphite and diamond are experimentally known as thermodynamically stable solids (graphite is thermodynamically stable at ambient condition and diamond under high pressure), although numerous metastable phases are known. For example, by applying pressure to graphite at low temperatures, a new superhard carbon allotrope was found, and its properties match those of one of the theoretically predicted structures (M-carbon)[8-11]. For the Ca-C system, the well-known Ca carbides include $CaC_2$ and $CaC_6$, whose high-pressure behaviors have been studied experimentally[12-14] and by *ab initio* calculations[15-19]. An interesting structural evolution has been uncovered under pressure: carbon atoms polymerize from dumbbells to 1D-chains to ribbons to 2D graphene sheets in compressed $CaC_2$[15] and from graphite sheets to a mixed $sp^2$-$sp^3$ structure in $CaC_6$[19]. Also, superconductivity was predicted in metallic high-pressure phases of $CaC_2$ with critical temperatures comparable to those observed in $CaC_6$[15].

Here, using variable-composition structure prediction code USPEX[1,8,20] the pressure-composition phase diagram of the Ca-C system was explored in order to fully understand the structural diversity and evolution of the C-C bonding types under high pressure. This resulted in five newly predicted stable stoichiometries ($Ca_5C_2$, $Ca_2C$, $Ca_3C_2$, $CaC$, and $Ca_2C_3$) with diverse carbon arrangements: isolated atoms in $Ca_2C$, hitherto unknown zigzag tetramers in $CaC$, and ribbons consisting of five-membered rings in $CaC_2$. Two phases ($Ca_2C$ and $Ca_2C_3$) were confirmed experimentally via *in situ* synchrotron powder X-ray diffraction (PXRD) measurements. Most surprising is that the low-pressure phase (monoclinic *C*2/*m* structure) of $Ca_2C$ exhibits quasi-two-dimensional metallic behavior and contains negatively charged calcium atoms. Also, strong interstitial electron localization was found in the newly predicted *R*-3*m* phase of $Ca_5C_2$, just as in compressed elements Li[21], Na[22] and Ca[5], as well as in the compound $Mg_3O_2$[2].

## Results

**Convex hull**. We have used the *ab initio* evolutionary algorithm USPEX[1,8,20], which can simultaneously find stable stoichiometries and the corresponding structures in multicomponent systems, to explore stable Ca-C compounds and their structures. In these calculations, all stoichiometries were allowed (with the constraint that the total number of atoms in the unit cell be below 16 atoms), and calculations were performed at 10 GPa, 20 GPa, 40 GPa, 80 GPa, and 100 GPa. The pressure-composition phase diagram of the Ca-C system is given in Fig. 1 a, in which the convex hull was obtained from the calculated enthalpies of the most stable structures for each composition at a given pressure. Thermodynamically, the convex hull at a given pressure connects the phases that are stable against decomposition into other binaries or the elements.

**Thermodynamically stable or metastable phases.** Using variable-composition evolutionary searches, we found that $Ca_5C_2$, $Ca_2C$, $Ca_3C_2$, CaC, $Ca_2C_3$, and $CaC_2$ have thermodynamic stability fields on the phase diagram: $Ca_2C_3$, stable from 0 to 28 GPa; $Ca_5C_2$, stable above 58 GPa; $Ca_2C$, stable above 14 GPa; $Ca_3C_2$, stable from ~50 GPa; CaC, stable above 26 GPa; and $CaC_2$, stable above 21 GPa (see Fig. 1 b and Fig. 2). For all the newly predicted structures, calculated phonon dispersion relations confirmed their dynamical stability. Surprisingly, our theoretical calculations show that the known phases of $CaC_2$ and $CaC_6$ are thermodynamically metastable at normal conditions (see Fig. 1 a); $CaC_2$ is thermodynamically stable only above 21 GPa, and $CaC_6$ does not have a thermodynamic stability field ($BaC_6$ is thermodynamically stable in the Ba-C system at 1 atm.[23]). We also explored metastable phases of $Ca_2C$ and CaC at lower pressure. The most stable low-pressure phase obtained for $Ca_2C$ has *C*2/*m* symmetry and that of CaC has *Immm* symmetry. The dynamical stability of these two thermodynamically metastable phases was confirmed via phonon calculations.

In order to analyze these predicted structures, we recall that the C-C bond length depends on the bond order and at 1 atm these lengths are 1.20 Å for the triple C-C bond, 1.33 Å for double bond, and 1.54 Å for single C-C bond. The carbon patterns predicted for calcium carbides, based on calculations presented in this work, are plotted in Figure 3. Combining this knowledge with the results of Bader analysis, we unravel very diverse chemistry. From the results of Bader analysis, one can clearly see the correlation between the charge and volume: negatively charged calcium atoms occupy significantly greater volume. Also we observe the decrease of C-C bond order from triple to double to single bonds as pressure increases. Note, however, that at pressures up to 100 GPa, the most carbon-rich stable compound is $CaC_2$. We consider the predicted phases in order of increasing carbon content.

**$Ca_5C_2$.** The stable structure of $Ca_5C_2$ has *R-3m* symmetry. It is a semimetal and is thermodynamically stable at pressures ranging from 58 GPa to at least 100 GPa. This phase has novel structural features: it can be described as consisting of alternating $CaC_2$ layers (where Ca is octahedrally coordinated by C atoms) and layers with composition $Ca_4$. The electron localization function (ELF) distribution in $Ca_5C_2$ shows strong charge transfer from Ca to C. Non-nuclear charge density maxima are located in the $Ca_4$ layer as plotted in Fig. 5 (ELF=0.75 at 58 GPa). Bader charges are +1.039 for Ca1, +0.823 for Ca2, +0.973 for Ca3, and -0.459 for the interstitial electron density maximum.

**$Ca_2C$.** Known alkali earth methanides include well-know $Be_2C$ (*Fm-3m*, Z=4) and $Mg_2C$ (antifluorite) recently synthesized by Kurakevych *et al.*[24]. However, no theoretical or experimental information has been reported on the methanide $Ca_2C$[25]. According to our calculations, $Ca_2C$ is thermodynamically stable above 15 GPa (space group *Pnma* (Z=4)). For *Pnma*-$Ca_2C$ we observe the largest negative charge of carbon atoms among all these phases: -2.321. In this semiconducting phase with band gap of 0.64 eV at 14 GPa, C atoms are isolated and one can represent this compound as a carbide with an idealized charge transfer scheme $(Ca^{2+})_2C^{4-}$ adhering to the Zintl concept. Metallic metastable $C2/m$-$Ca_2C$ has a unique structure, consisting of alternating layers of stoichiometry $Ca_2(C_2)$ and $Ca_2$ (two kinds of calcium atoms play distinctly different roles, see Fig. 6 a), and these layers have net charges of +0.582 and -0.582, respectively. What is unusual is that the Ca-layer is negatively charged, *i.e.*, it is a reservoir of electrons. To further analyze this phenomenon, it is instructive to look first at the $Ca_2(C_2)$ layer. This $C_2$-group can be represented as having a triple C-C bond and its ideal charge is -2 (Bader charge is -1.892), and if each Ca had the ideal charge of +2, the total charge of the $Ca_2(C_2)$ layer would be +2, and two electrons would be transferred to the $Ca_2$ layer. In reality, the C-C bond here has a somewhat lower order (C-C distance is 1.28 Å at 5 GPa) and therefore takes more electrons from Ca atoms, leaving less for the $Ca_2$ layer, but not changing the picture qualitatively. ***To our knowledge, this is the first example of negatively charged metal atoms in a compound with more electronegative atoms.*** Note the enormous difference of Bader volumes of the positively and negatively charged Ca atoms (16.570 vs 41.901 Å$^3$). One can expect that the electrons in the Ca-layer are very loosely bound and the work function of this compound can be expected to be extremely low. The density of states of metastable $C2/m$ phase of $Ca_2C$ reveals a remarkable step-like feature near the bottom of the valence band, followed by a nearly constant density of states (see Fig. 6 b), presenting an example of a quasi-two-dimensional electronic structure as observed in Li-Be alloys[26]. The calculated Fermi surface of $C2/m$-$Ca_2C$ at 3 GPa holds a hollow square cylinder-like Fermi surface along the $\Gamma$-V direction (*i.e.*, reciprocal lattice basis vector $\boldsymbol{b}_3$ direction) in the Brillouin zone, signaling quasi-two-dimensional electronic properties (see Fig. 6 c).

**$Ca_3C_2$.** For $Ca_3C_2$, no thermodynamically stable phase exists below 50 GPa. A metastable *P4/mbm* (Z=2) phase, favored in the pressure range from 5 GPa to 30 GPa, transforms into a *C2/c* (Z=4) structure at 30 GPa, which is thermodynamically stable above 50 GPa. The structure of *P4/mbm*-$Ca_3C_2$ contains doubly-bonded $C_2$-groups (C-C distance 1.39 Å at 20 GPa), with an ideal charge -4, *i.e.*, accepting four electrons from calcium atoms, leaving two electrons per formula to form Ca-Ca bonds in this metallic compound. Metallic *C2/c*-$Ca_3C_2$ with a pseudogap at the Fermi level has singly-bonded $C_2$ groups (C-C bong length 1.51 Å at 38.7 GPa), which have ideal charge -6, exactly balanced by three Ca atoms in the formula.

**CaC.** Metallic CaC has two thermodynamically stable phases below 100 GPa. At 14 GPa, the metastable orthorhombic *Immm* structure transforms into a monoclinic $P2_1/c$ structure (stable thermodynamically above 26 GPa, favored over a wide pressure range of 14-57.5 GPa), followed by a thermodynamic stable *Imma* structure. $P2_1/c$-CaC is very interesting because its structural formula $Ca_4(C_4)$ contains a unique and hitherto unknown zigzag $C_4$- group, with C-C distances between 1.48-1.50 Å at 14 GPa, indicating bond orders between 1 and 2 and ideal charges of about -2.5 for the end C atoms (Bader charge -1.447) and about -1 for the central C atoms (Bader charge -0.905). *Imma*-CaC has infinite zigzag chains of C atoms (C-C bond length of 1.55 Å at 57.5 GPa, indicating a weakened single bond) in the y-axis direction. The structural formula of metastable *Immm*-CaC is $Ca_2(C_2)$, and with a doubly bonded $C_2$-group (C-C distance 1.33 Å at 7.1 GPa) that has an ideal charge of -4 (Bader charge –2.340), it exactly balances the ideal charge of two Ca atoms. All three phases of CaC beautifully conform to the trend of increasing polymerization of the C-sublattice with increasing pressure.

**$Ca_2C_3$.** The structure of $Mg_2C_3$ (space group *Pnnm*, Z = 2), the only known alkaline-earth metal allylenide with $C_3^{4-}$ anions[25], was considered when searching for stable phases of $Ca_2C_3$. Total energy calculations exclude the possibility of the ambient-pressure $Mg_2C_3$-type structure. The semiconducting *C*2/*m* structure (band gap of 1.06 eV at 10 GPa) is instead the most stable one below 34.5 GPa (thermodynamically stable from 0-28 GPa), followed by metastable *C*2/*c* structure. By fitting energy vs. volume data to the 3$^{rd}$-order Birch-Murnaghan equation of state[27], the calculated bulk modulus $B_0$ of *C*2/*m*-$Ca_2C_3$ is about 89 GPa, which is higher than that of $CaC_2$ (50 GPa). At ~40 GPa, the metallic *C*2/*c* structure transforms into a metastable *P*-1 structure (metal), which dominates the pressure range between 40-65 GPa. At higher pressures, a metallic metastable *Imma* structure is stable and contains zigzag carbon chains (Figs. 3 and 4). We searched at much higher pressures for 3D-polymeric carbon frameworks in $Ca_2C_3$, but found none at pressures up to at least 300 GPa. For comparison, in $CaC_2$ we have found that graphene sheets predicted in the high-pressure phase can be stable up to at least 1 TPa[15].

For $Ca_2C_3$, the carbon arrangement changes from isolated $C_3$ to carbon chains to ribbons (Fig. 3). The structure of *C*2/*m*-$Ca_2C_3$ can be described as $Ca_2$ layers linked together by nearly linear symmetric $C_3$ groups with double C-C bonds (C-C distances 1.32 Å at 18.1 GPa). With this configuration, the total charge of the $C_3$ group should be -4 (Bader charge -2.692), exactly compensating the charge of two Ca atoms in the formula. Central carbon atoms in the $C_3$ group in this valence scheme should be neutral, and yet turn out to have the largest negative Bader charge of -0.738, whereas the end atoms, whose idealized charge is -2, develop a lower Bader charge (-0.977). This discrepancy is explained by the effects of Ca atoms, which form significant bonds with the central carbon atom in the $C_3$ group and transfer some electronic charge to them. Most recently, some of us reported the prediction and synthesis of β-$Mg_2C_3$[28], which is isostructural with our *C*2/*m*-$Ca_2C_3$ reported here. The structure of *C*2/*c*-$Ca_2C_3$ (C-C distances 1.43-1.47 Å at 34.5 GPa) has an idealized charge transfer scheme $Ca^{4+}_2C^{4-}_3$. In this metallic phase, C-atoms are polymerized into infinite chains with nearly closed six-member rings running through channels of Ca host lattice. *P*-1-$Ca_2C_3$ features a complicated extended 1D-ribbon of carbon atoms with nearly single C-C bonds (lengths 1.47-1.50 Å at 40 GPa).

*Imma*-$Ca_2C_3$ has a very interesting structure with extended 1D-ribbons of carbon atoms cut from the graphene layer. Bond lengths in this ribbon are 1.50-1.52 Å at 70 GPa, slightly longer than in graphene and indicating predominantly single bonds. Electronic structure calculations show that both *P*-1 and *Imma* phases of $Ca_2C_3$ are metals. Based on Allen and Dynes modified equation[29], we have checked for superconductivity in these phases at 34 GPa and 65 GPa, respectively, and found none.

**CaC$_2$.** CaC$_2$ is thermodynamically stable above 21 GPa (see Fig. 1 a). The lower pressure phases *C*2*/m* and *Cmcm* reported previously[15] are metastable, which could be unraveled by calculating enthalpy of formation ($\Delta H_f$) at lower pressure. Considering that graphite is the ground state of carbon at zero pressure, we performed additional calculations where the van der Waals (vdW) interaction is accounted for by using the optPBE-vdW functional[30] in our calculation. At zero pressure, the calculated $\Delta H_f$ per atom (-0.17 eV) of CaC$_2$ is close to standard $\Delta H_f$ per atom (-0.21 eV at 298K and 1 atm[31]) but higher than that (-0.27 eV) of Ca$_2$C$_3$, confirming the thermodynamic metastability of CaC$_2$ at ambient conditions (see Fig. 1 a). It is very unexpected, but the above numbers fully confirm this conclusion, that the well-known and industrially produced compound CaC$_2$ is metastable at ambient conditions, while the so far never seen compound Ca$_2$C$_3$ is actually stable. This could be either due to kinetics, or due to conditions of synthesis. In addition to our previous result[15], we found a new phase with *P*-1 symmetry, which contains infinite carbon chains with five-membered rings (C-C distance is between 1.442-1.507 Å at 20 GPa (see Fig. 3 i), signaling single or double bonds), and is the lowest enthalpy structure over a wide pressure range from 7.5 GPa to 37 GPa (thermodynamically stable from 21 GPa to 37 GPa, see Fig. 4). With further application of pressure, metallic *P*-1-CaC$_2$ transforms into metallic *Immm*-CaC$_2$[15], in which carbon atoms are polymerized to form quasi-one dimensional ribbons (see Figs. 2-3).

**Experiments.** In order to confirm theoretical structure search predictions, we performed synthesis studies under high-pressure / high-temperature conditions. Diamond anvil cells were loaded with both calcium- and carbon-rich Ca+C mixtures, compressed to pressures up to 25 GPa, heated to temperatures up to approximately 2000 K and probed *in situ* using synchrotron PXRD. Under these pressure conditions, the formation of *Immm*-$CaC_2$, *C2/m*-$Ca_2C_3$ and *Pnma*-$Ca_2C$ may be expected based on thermodynamic stabilities, as these phases are the only stable ones that appear on the convex hull up to 25 GPa (see Fig. 1); indeed, two of these three structures were observed experimentally.

When samples were compressed above ~10 GPa and heated to ~2000 K, mixtures of elemental glassy carbon and face-centered cubic (*fcc*) Ca transformed into a new low-symmetry phase. After comparison with density functional theory (DFT) structure predictions, PXRD reflections originating from this phase were readily indexed to the monoclinic *C2/m*-$Ca_2C_3$ structure with excellent agreement between experiment and theory. Figure 7 shows experimental PXRD data obtained at 17 GPa with $a$ = 5.169(4) Å, $b$ = 4.994(3) Å, $c$ = 6.322(3) Å and $\beta$ = 128.53(3)° which compares with $a$ = 5.151 Å, $b$ = 4.962 Å, $c$ = 6.306 Å and $\beta$ = 128.81° for calculations relaxed at 18 GPa. This sample was decompressed in steps of ~2 GPa to obtain lattice parameters as a function of pressure (see Fig. 8). Theoretical lattice parameters show an average absolute deviation (AAD) 0.3% from experiment over the entire pressure range, and the *C2/m*-$Ca_2C_3$ phase was recoverable to ambient pressure, but was found to be air / moisture sensitive. The experimental *P-V* data were fit to a second-order Birch-Murnaghan equation of state with $B_0$ = 84(2) GPa, in good agreement with theoretical predictions (89 GPa).

At pressures above ~22 GPa, a second carbide phase (*Pnma*-$Ca_2C$) was synthesized upon laser heating. This same phase was reproducibly formed both from elemental Ca+C mixtures and from samples containing *C*2/*m*-$Ca_2C_3$, indicating disproportionation of $Ca_2C_3$ into a more stable carbide phase above ~22 GPa. Figure 7 shows experimental PXRD data at 24 GPa with $a$ = 6.122(1) Å, $b$ = 4.004(1) Å, $c$ = 7.223(1) Å, which compares with $a$ = 6.044 Å, $b$ = 3.977 Å, $c$ = 7.265 Å for DFT calculations at the same pressure. Calculated lattice parameters show an AAD of 0.5% from experimental values between 25-5 GPa (see Fig. 8), which was the lowest pressure obtained due to failure of a diamond anvil. Fitting the P-V data to a second-order Birch-Murnaghan equation of state yields $B_0$ = 53(4) GPa.

## Discussion

We find that the carbon sublattice within all predicted carbide phases has close correlation with the Ca:C ratio (see Fig. 2). With increasing carbon content, isolated carbon atoms are polymerized, in turn, into $C_2$ dumbbells, $C_3$ and $C_4$ groups, chains, ribbons and graphene sheets (see Figs. 4). The polymeric carbon structures reveal an expected trend when comparing to the structural chemistry of the heavier congeners of the 4$^{th}$ main group elements in Zintl phases (alkali or alkaline-earth silicides, germanides, and stannides) [16, 32]. Yet in spite of certain similarities to silicides, calcium carbides differ from them due to distinct bonding features. Combining present analysis and our previous results [15,16,19], one can conclude that for the Ca-C system, one can cover *sp* to *sp*$^2$ to *sp*$^2$+*sp*$^3$ to *sp*$^3$ hybridizations of carbon as pressure increases. This pressure-induced structural evolution of carbon was also found in other alkali metal or alkaline-earth metal carbides[16,19,23]. Together with our previous results for $CaC_2$[15] and $CaC_6$[19], it is clear that a three-dimensional network of carbon in $CaC_x$ can be formed when $x$ is greater than 2, (from sheets to three-dimensional frameworks to Ca-C phase separation with slabs of diamond at high C content), consistent also with the behavior of the metastable $CaC_4$ compound found in our structural searches. On the other hand, the structural features of carbon-rich compounds[19] can be extended to alkali-metal or alkaline-earth metal congeners of the group four elements, which allows one to fabricate a variety of the 3D framework structures of the group four elements by removing metal sublattices. The unexpected mechanical[19,33] or electronic characteristics[34] uncovered in these 3D framework structures pave the way to novel materials.

In summary, we have produced the first complete pressure-composition phase diagram for $CaC_x$ compounds at pressures up to 100 GPa and demonstrated the experimental synthesis of two previously unknown compounds ($Ca_2C_3$ and $Ca_2C$), validating part of our predicted phase diagram. Contrary to normal ionic compounds, there is no "dominant" compound stable in this whole pressure range. The well-known $CaC_2$ and $CaC_6$ were found to be metastable at normal conditions; $CaC_2$ is stable only above 21 GPa, and $CaC_6$ is never thermodynamically stable, while hitherto unreported $Ca_2C_3$ is actually thermodynamically stable at ambient pressure. Bader analysis unravels very diverse chemistry: the decrease of C-C bond order from triple to double to single bonds at increasing pressure; a negatively charged metal layer in calcium-rich $Ca_2C$ compound; a hitherto unknown bent linear $C_4$ group in the $P2_1/c$ phase of CaC; C-ribbons being present in carbon-rich compounds. The $C2/m$-$Ca_2C$ phase provides a fresh and very attractive example of a two-dimensional metal, presenting the only known example of a compound where a metal atom (Ca) develops a negative Bader charge in presence of a more electronegative atom (C). Such unusual compounds are likely to find potential applications if synthesized in sufficiently large quantities. While powerful computational methods, such as USPEX, are capable of reliably predicting these exotic compounds, simple theoretical models capable of anticipating them are yet to be developed.

## Methods

**Structure search and theoretical calculations.** Searches for stable structures of the Ca-C system under compression were carried out using the evolutionary algorithm USPEX in combination with the VASP code[35] based on density functional theory within the generalized gradient approximation with the exchange-correlation functional of Perdew, Burke and Ernzerhof[36], employing the projector augmented-wave (PAW)[37] method where $2s^22p^2$ and $3s^23p^64s^2$ are treated as valence electrons for C and Ca atoms, respectively. For carbon, a 'hard' pseudopotential was used in searching for stable structures. For the crystal structure searches, we used a plane-wave basis set cutoff of 700 eV and performed the Brillouin zone integrations using a coarse k-point grid. The most interesting structures were further relaxed at a higher level of accuracy with a basis set cutoff of 1000 eV and a k-point grid of spacing $2\pi \times 0.018$ Å$^{-1}$. Iterative relaxation of atomic positions was stopped when all forces were smaller than 0.001 eV Å$^{-1}$. For compounds predicted via variable-composition searches, we re-searched their stable structures using fixed-composition calculations, with two, three, and four formula units per unit cell. For $Ca_2C_3$, some evolutionary calculations were also performed under the pressure of 30 GPa, 50 GPa, 80 GPa, 120 GPa, 160 GPa, 240 GPa, and 300 GPa with two or four chemical formula units per unit cell so as to discern the possibility of three-dimensional network carbon.

The enthalpy of formation per atom of $Ca_nC_m$ is defined as $\Delta H_f(Ca_nC_m)=[H(Ca_nC_m)-nH(Ca)-mH(C)]/(n+m)$, where all enthalpies H are given at the same pressure and zero temperature. At a given pressure, the calcium carbides located on the convex hull are thermodynamically stable against decomposition to any other binaries or the elements while the compound above the convex hull meta-stable.

Bader analysis was performed for exploring chemical bonding and local electrons[38]. To get a converged charge density, the plane wave kinetic energy cutoff of 1000 eV and Monkhorst-Pack **k**-point meshes with the reciprocal space resolution of $2\pi\times0.02$ Å$^{-1}$ were used for all the structures. A series of FFT grids to accurately reproduce the correct total core charge were tested by increasing parameters NG(X,Y,Z)F to 1.5, 2 and 2.5 times the default one.

The lattice dynamics and superconducting properties of $Ca_2C_3$ were calculated by the Quantum ESPRESSO package[39] using Vanderbilt-type ultrasoft pseudopotentials with cutoff energies of 50 Ry and 500 Ry for the wave functions and the charge density, respectively. The electronic Brillouin zone (BZ) integration in the phonon calculation was based on a $12\times12\times12$ of Monkhorst-Pack k-point meshes. The dynamic matrix was computed based on a $6\times6\times6$ mesh of phonon wave vectors. The electron-phonon coupling was convergent with a finer grid of $48\times48\times48$ **k** points and a Gaussian smearing of 0.01 Ry. For other compounds, phonon calculations were performed using the Phonopy code[40]. The Fermi surface of $C2/m$-$Ca_2C$ at 3 GPa was calculated via Quantum Espresso code using Vanderbilt-type ultrasoft pseudopotentials with $16\times16\times8$ of Monkhorst-Pack k-point meshes.

**Experiment.** Reagents for experimental studies consisted of commercial calcium metal (Sigma-Aldrich, dendritic pieces, 99.99%) and glassy carbon (Sigma-Aldrich, 2-12 μm, 99.95%). The carbon powder was degasified for 12 hours at ~200°C in a vacuum oven and then sealed under Ar. Diamond anvil cell syntheses (up to 25 GPa and 2000 K with *in situ* PXRD) were performed at HPCAT, sector 16, of the Advanced photon Source. A small amount of carbon powder and Ca metal shavings were pressed in thin layers within a rhenium gasket in a diamond anvil cell equipped with 400-μm culet diamonds inside of an inert Ar glovebox ($O_2$ < 1 ppm; $H_2O$ < 1 ppm). It was not possible to control the precise Ca:C ratio, but compositions were estimated to range between $0.333 \leq$ Ca:C $\leq 3$, based upon the volume of material loaded into the diamond cell. Samples were sealed inside the glovebox without pressure medium or loaded with Ne to improve thermal isolation from diamonds and quasi-hydrostatic conditions for subsequent diffraction measurements. Pressure was estimated using the Ne equation of state[41] and/or with a ruby gauge[42]. To generate high temperatures, samples were heated on both sides using IR fiber lasers. In some cases two or three heating cycles were performed, with X-ray diffraction patterns collected before, during, and after laser heating. During laser heating, temperatures from each side of the sample were estimated by collecting emitted thermal radiation, correcting for the optical system response and fitting the spectral data to Planck's equation. Angle-dispersive PXRD patterns were collected using a MAR image plate calibrated with a $CeO_2$ standard. Patterns were integrated using FIT2D[43] and phase recognition and indexing were performed using PowderCell and CheckCell programs[44]. While observed PXRD intensities were in good agreement with DFT-derived structural models, observed powder statistics were not suitable for Rietveld refinements. Full profile refinements were performed using the Le Bail intensity extraction method, as implemented in GSAS[45] with EXPGUI[46].

## Acknowledgements

Y.L.L acknowledges support from the NSFC (11347007), Qing Lan Project, and the Priority Academic Program Development of Jiangsu Higher Education Institutions (PAPD). A.R.O and S.N.W thank the National Science Foundation (EAR-1114313, DMR-1231586), DARPA (Grants No. W31P4Q1210008 and No. W31P4Q1310005), the Government of Russian Federation (No. 14.A12.31.0003) for financial support, and Foreign Talents Introduction and Academic Exchange Program (No. B08040). Calculations were performed on XSEDE facilities and on the cluster of the Center for Functional Nanomaterials, Brookhaven National Laboratory, which is supported by the DOE-BES under contract no. DE-AC02-98CH10086. Experimental efforts were supported by DARPA under grant No. W31P4Q1310005. Portions of this work were performed at HPCAT (Sector 16), Advanced Photon Source (APS), Argonne National Laboratory. HPCAT operations are supported by DOE-NNSA under Award No. DE-NA0001974 and DOE-BES under Award No. DE-FG02-99ER45775, with partial instrumentation funding by NSF. The Advanced Photon Source is a U.S. Department of Energy (DOE) Office of Science User Facility operated for the DOE Office of Science by Argonne National Laboratory under Contract No. DE-AC02-06CH11357.

## Author contributions

Y.L.L and A.R.O designed the research. Y.L.L, A.R.O, and S.N.W performed calculations. H.G., J.S.S and T.A.S performed the experimental studies. All authors analyzed the data and contributed to write the paper.

**Figure legends**

**Figure 1. Stability of new calcium carbides.** (**a**) Convex hull diagram for the Ca-C system at selected pressures. At a given pressure, the compounds located on the convex hull are thermodynamic stable. (**b**) Pressure-composition phase diagram of the Ca-C system. Thick solid lines represent thermodynamically stable phases and dashed lines represent metastble phases (Red lines represent metal and blue semiconductor.).

**Figure 2. The predicted crystal structures of stable Ca-C compounds.** (**a**) Thermodynamically stable *Pnma* structure of $Ca_2C$. (**b**) The metastable low pressure *P4/mbm* structure of $Ca_3C_2$. (**c**) Thermodynamiclly stable high pressure *C2/c* structure of $Ca_3C_2$. (**d**) Thermodynamically stable $P2_1/c$ structure of CaC. (**e**) Thermodynamically stable high pressure *Imma* structure of CaC. (**f**) $Ca_2C_3$ crystallizes in *C2/m* structure at pressures up to 28 GPa. (**g**) Thermodynamically metastable *C2/c* structure of $Ca_2C_3$. (**h**) Thermodynamically metastable *P*-1 structure of $Ca_2C_3$. (**i**) Thermodynamically metastable high pressure *Imma* of $Ca_2C_3$. (**j**) Thermodynamically stable *P*-1 structure of $CaC_2$. (**k**) Thermodynamically stable *Immm* structure of $CaC_2$. The blue and brown spheres represent calcium and carbon atoms, respectively.

**Figure 3. Carbon patterns in the Ca-C system.** (**a**) The isolated carbon anions in the *Pnma* structure of $Ca_2C$. (**b**) Cdimers observed in the *C2/m* structure of $Ca_2C$, *P4/mbm* and *C2/c* structures of $Ca_3C_2$, *Imma* structure of CaC, *C2/c* structures of $Ca_2C_3$, and *C2/m* or *C2/c* structures of $CaC_2$. (**c**) The carbon trimer occurs in the *C2/m* structure of $Ca_2C_3$ at zero pressure. (**d**) Zigzag $C_4$ groups observed in $P2_1/c$ structure of CaC at 20 GPa. (**e**) Zigzag carbon chains in the *Imma* structure of CaC at 58 GPa. (**f**) Carbon chains with two types of carbon-carbon bondings in *C2/c* structure of $Ca_2C_3$ at 34.5 GPa. (**g**) Carbon chains with four types of carbon-carbon bongdings in *P*-1 structure of $Ca_2C_3$ at 40 GPa. (**h**) Armchair carbon chains in the *Cmcm* structure of $CaC_2$ at 4 GPa. (**i**) Carbon stripes in the *P*-1 structure of $CaC_2$ at 20 GPa. (**j**) Carbon ribbons in the *Immm* structure of $CaC_2$ at 10 GPa. (**k**) Carbon ribbons in the *Imma* structure of $Ca_2C_3$ at 65 GPa. Bond lengths (in Å) are given. The inequivalent

C1, C2, C3, and C4 (occupying different Wyckoff positions in the unit cell) are shown by red, blue, yellow, and grey spheres, respectively.

**Figure 4. Carbon arrangement with increasing carbon content.** Only thermodynamically stable phases are shown. Additionally, the conducting properties are shown.

**Figure 5. Electron localization function (ELF) of *R-3m*-Ca$_5$C$_2$ at 60 GPa.** The isosurface ELF=0.75 is shown. The observed interstitial electron charge accumulation shows that Ca$_5$C$_2$ with *R-3m* symmetry is an electride.

**Figure 6. Structural and electronic property of *C2/m* phase of Ca$_2$C at 3 GPa. (a)** The top view of *C2/m* structure along the y axis shows a clearly layered structure. **(b)** Total electronic density of states (DOS). **(c)** Fermi surface of the *C2/m* structure. The indication of a quasi-two-dimensional metal in total DOS is confirmed by the hollow square prismatic cylinder-like Fermi surface.

**Figure 7. PXRD patterns for experimentally observed Ca-C phases. (a)** *Pnma*-Ca$_2$C synthesized at 24 GPa ($wR_p$ = 1.5%, $R_p$ = 0.9%). **(b)** *C2/m*-Ca$_2$C$_3$ synthesized at 17 GPa ($wR_p$ = 2.0%, $R_p$ = 0.9%). Experimental data (points) are compared with full-profile refinements using the Le Bail method (blue lines), with differences (red lines). Simulated powder patterns using atomic positions derived from DFT-optimized structures are shown for intensity compasion (green lines). Positions of reflections of Ca$_2$C (**a**) or Ca$_2$C$_3$ (**b**), Ca, Ne and a third carbide phase are indicated by black, green, blue and red tick marks, respectively.

**Figure 8. Lattice parameters and unit cell volumes for Ca-C phases.** Experimental parameters (points) are compared with DFT predictions (dashed lines) for both *C2/m*-Ca$_2$C$_3$ (**a**) and *Pnma*-Ca$_2$C (**b**). Experimentally-derived equations of state for both phases are shown as solid lines in the lower panels.

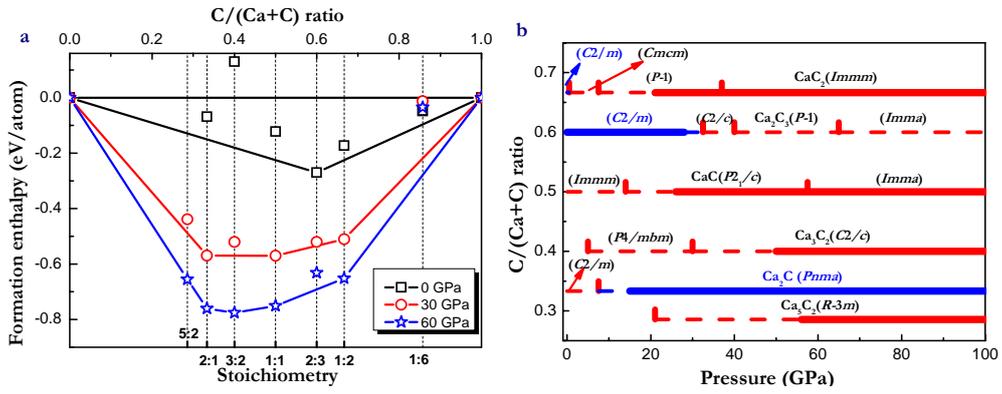

**Li   *et al*.   Fig. 1**.

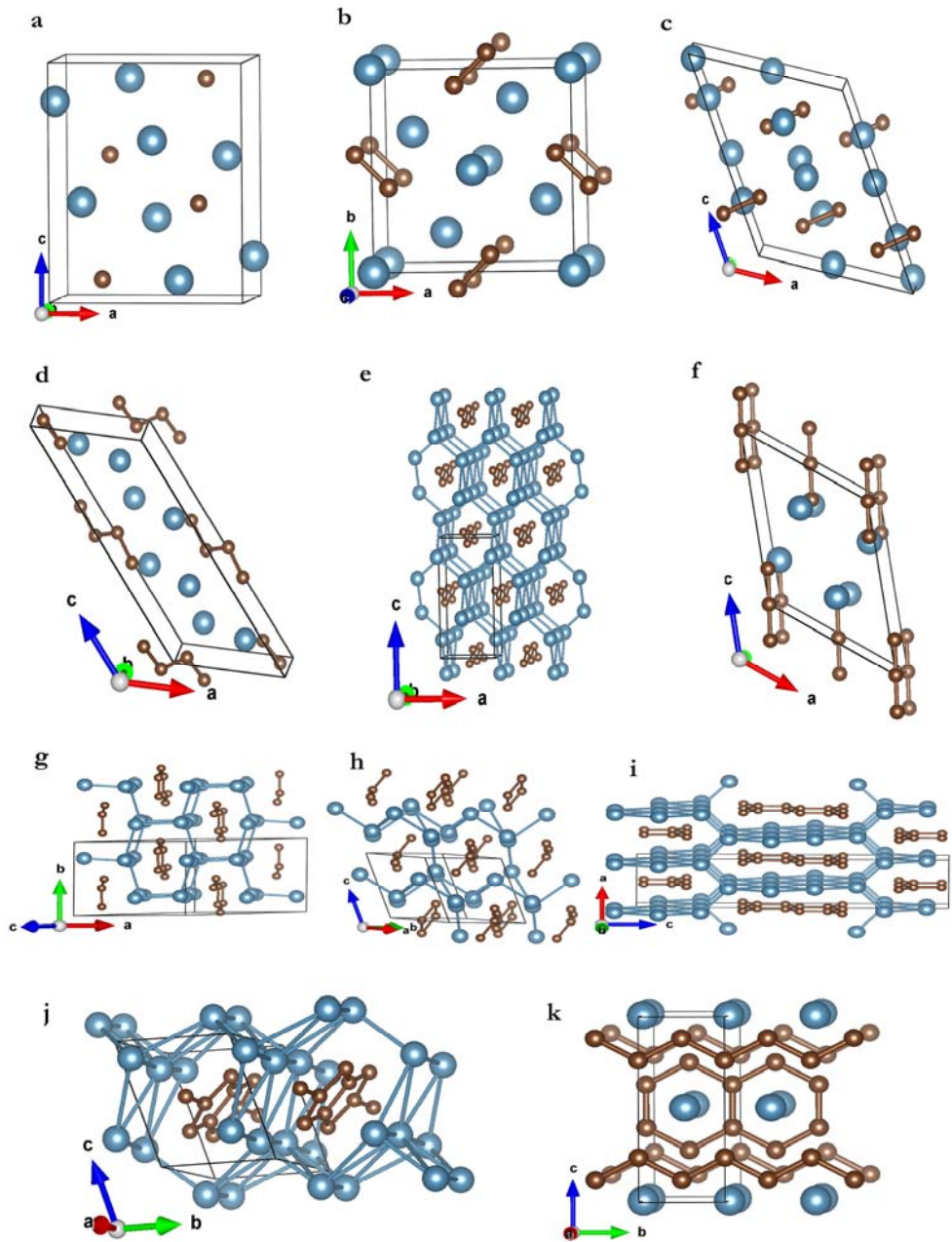

**Li** *et al.* **Fig. 2**.

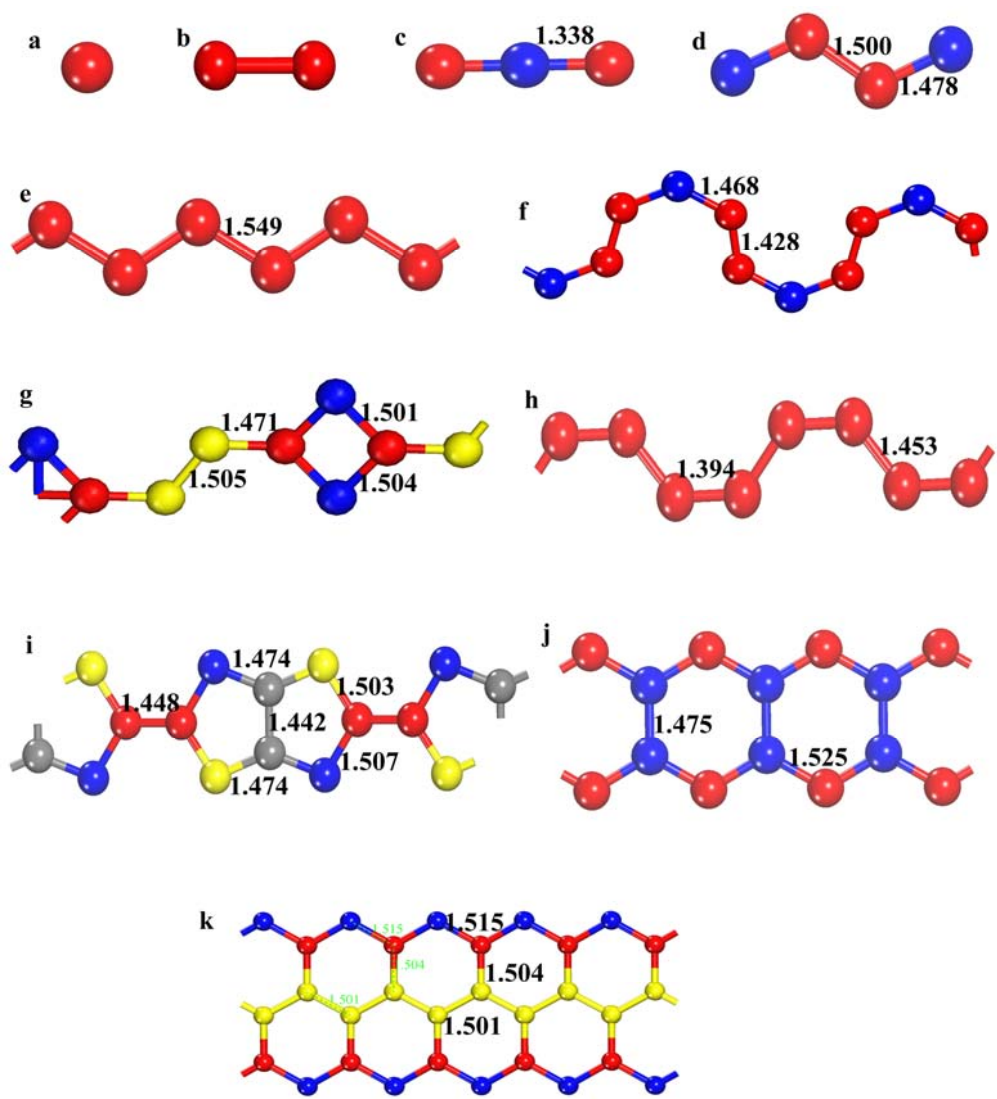

Li *et al.*   Fig. 3.

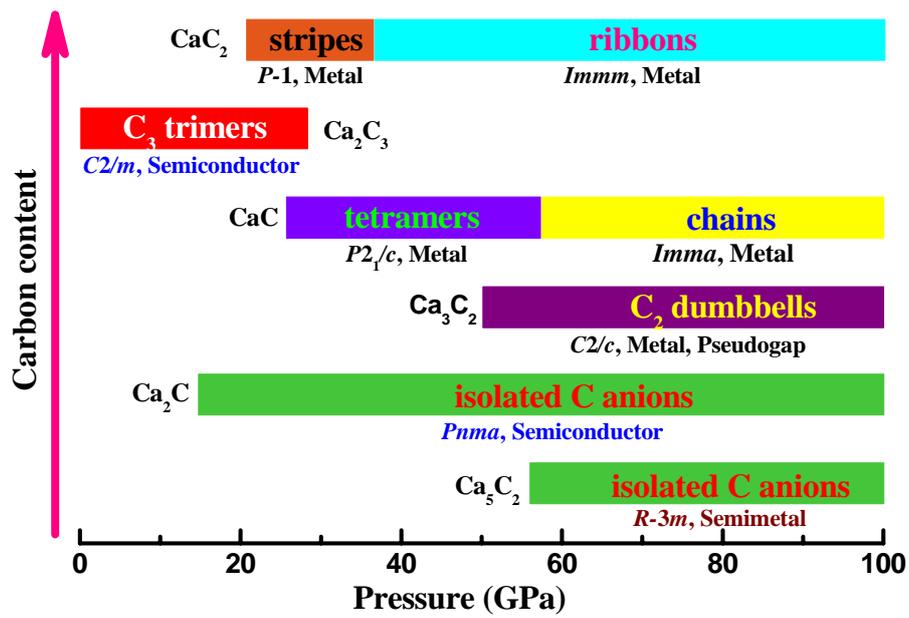

**Li** *et al.* **Fig. 4**.

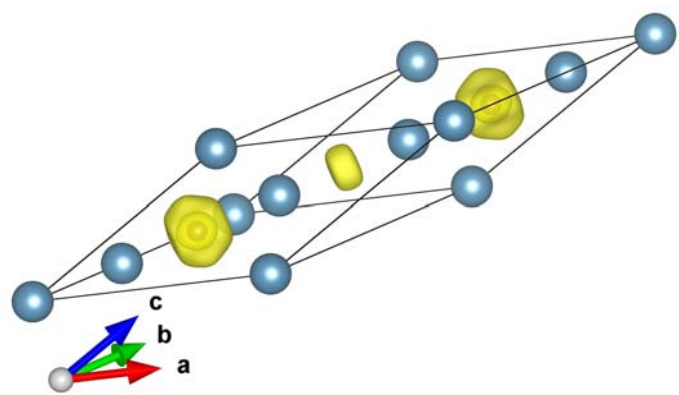

**Li** *et al*. **Fig. 5**.

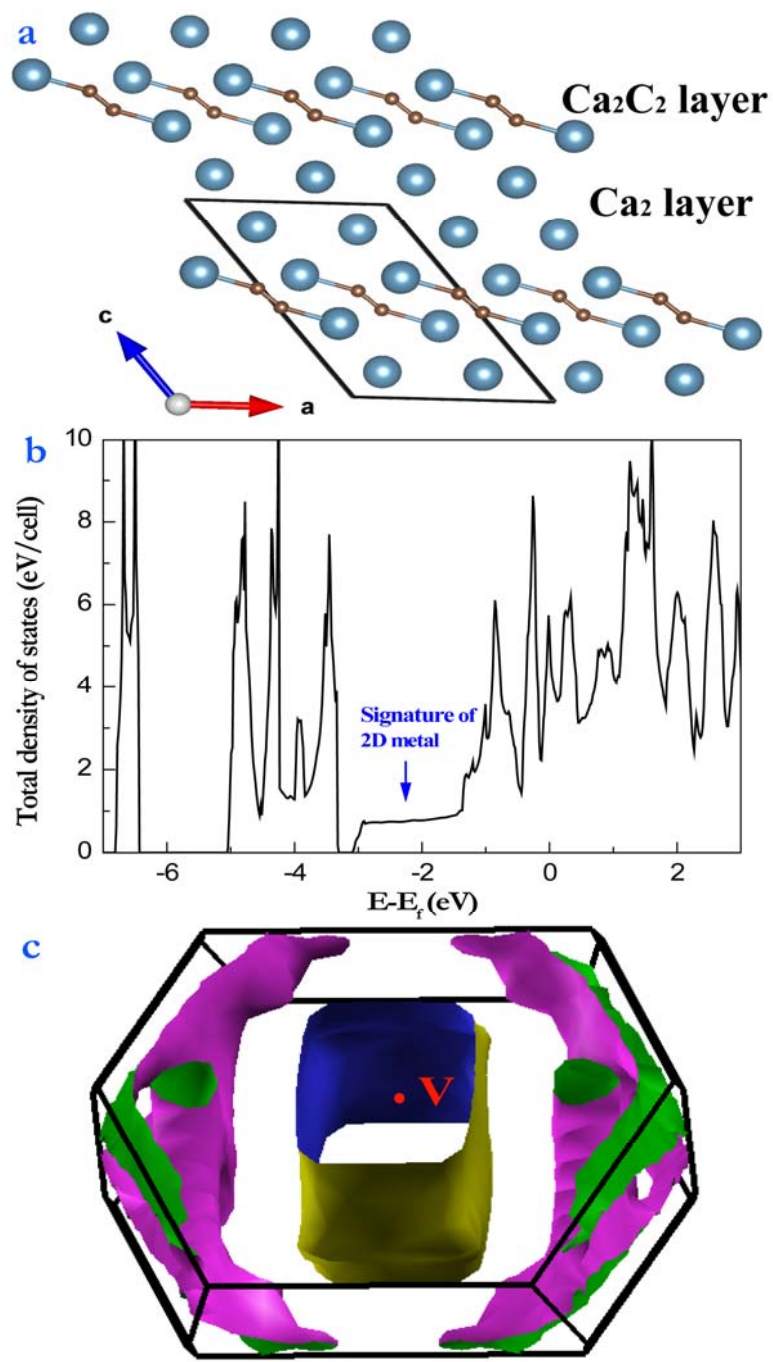

Li  *et al*.    Fig. 6.

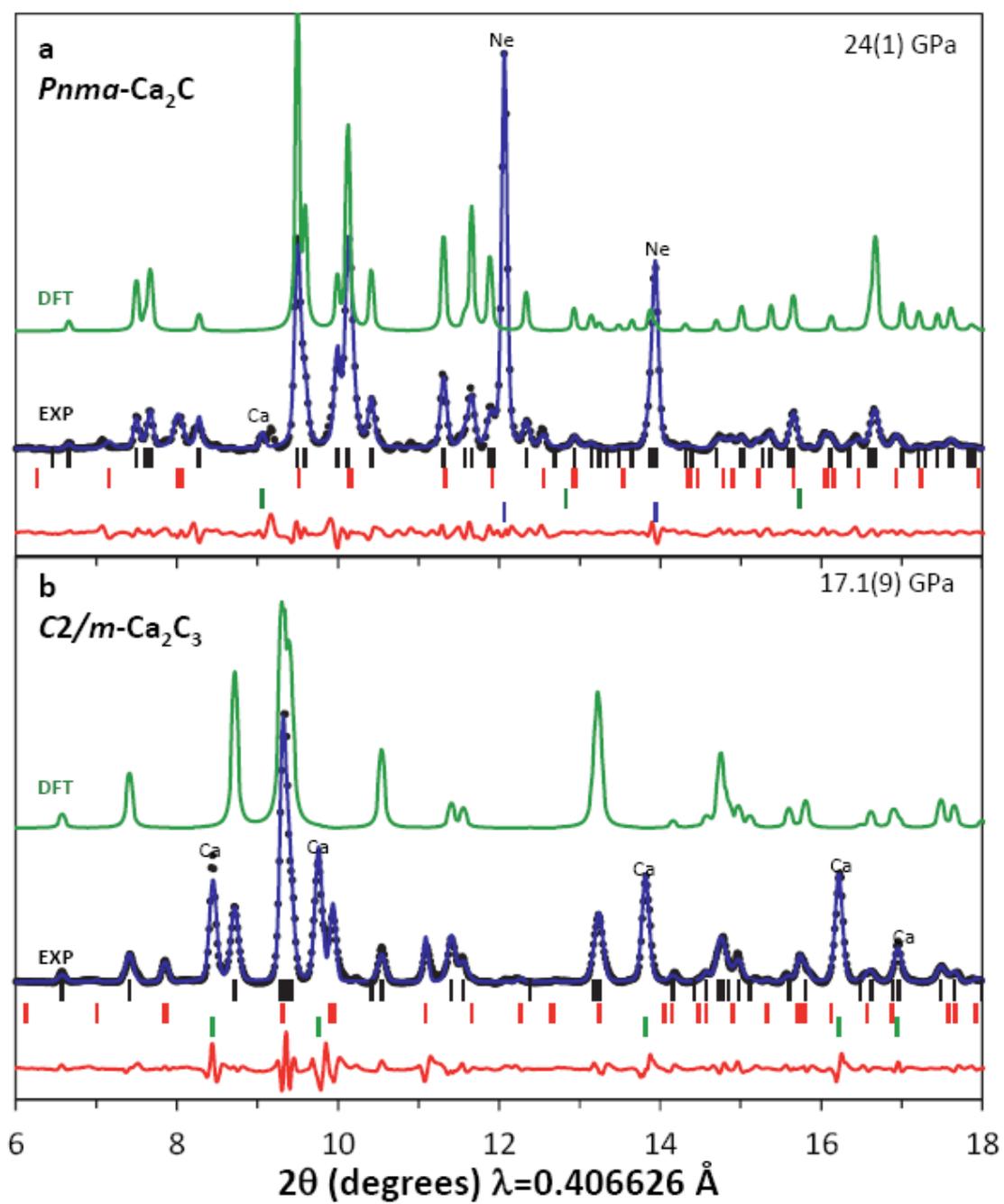

Li *et al.* Fig. 7

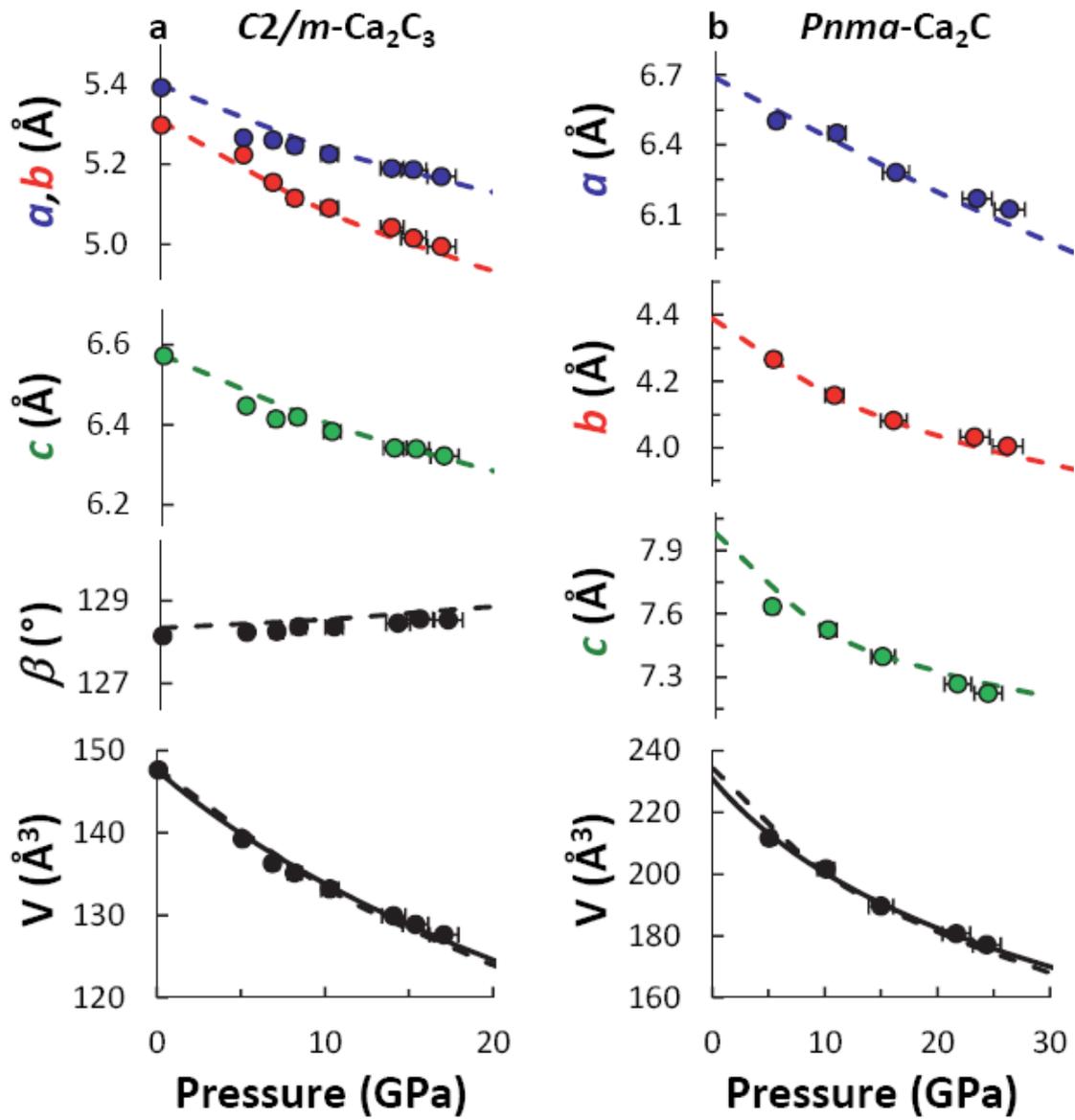

**Li** *et al.* **Fig. 8**